\documentclass[12pt]{article}
\newcommand{\bq}{\begin{equation}}
\newcommand{\eq}{\end{equation}}
\newcommand{\ba}{\begin{eqnarray}}
\newcommand{\ea}{\end{eqnarray}}

\newcommand{\nl }{ \nonumber  }

\newcommand{\p}{\partial}
\newcommand{\h}{\hspace{.5cm}}
\newcommand{\s}{\sigma}

\begin{document}
\vspace*{1.cm}
\begin{center}
{\bf INTEGRABLE SYSTEMS FROM MEMBRANES ON $AdS_4\times S^7$

\vspace*{1cm} P. Bozhilov}\footnote{e-mail: bozhilov@inrne.bas.bg}

\ \\
\textit{Institute for Nuclear Research and Nuclear Energy,\\
Bulgarian Academy of Sciences, \\ 1784 Sofia, Bulgaria}

\end{center}
\vspace*{.5cm}

We describe how Neumann and Neumann-Rosochatius type integrable
systems, as well as the continuous limit of the $SU(2)$ integrable
spin chain, can be obtained from membranes on $AdS_4\times S^7$
background, in the framework of AdS/CFT correspondence.


\vspace*{.5cm} {\bf PACS}:11.25.Mj

\vspace*{.5cm} {\bf Keywords:} M-theory, integrable systems,
AdS-CFT duality.

\section{Introduction}
\hspace{.5cm}The $2$-branes (membranes) and $5$-branes are the
fundamental dynamical objects in the eleven dimensional
$M$-theory, which is the strong coupling limit of the five
superstring theories in ten dimensions, and whose low energy field
theory limit is the eleven dimensional supergravity.

It is known that large class of classical string solutions in the
type IIB $AdS_5\times S^5$ background is related to the Neumann
and Neumann-Rosochatius integrable systems, including recently
discovered spiky strings and giant magnons \cite{NR2}. It is also
interesting if these integrable systems can be associated with
some membrane configurations in M-theory. We explain here how this
can be achieved by considering membrane embedding in $AdS_4\times
S^7$ solution of $M$-theory, with the desired properties.

On the other hand, we will show the existence of membrane
configurations in $AdS_4\times S^7$, which correspond to the
continuous limit of the $SU(2)$ integrable spin chain , arising in
$\mathcal{N}=4$ SYM in four dimensions, dual to strings in
$AdS_5\times S^5$ \cite{K0311}.

\section{Membranes on $AdS_4\times S^7$}
\h We start with the following membrane action \ba\label{oma} S&=&
\int d^{3}\xi\mathcal{L}\\ \nl &=& \int
d^{3}\xi\left\{\frac{1}{4\lambda^0}\Bigl[G_{00}-2\lambda^{j}G_{0j}+\lambda^{i}
\lambda^{j}G_{ij}-\left(2\lambda^0T_2\right)^2\det G_{ij}\Bigr] +
T_2 C_{012}\right\},\ea where \ba\nl &&G_{mn}= g_{MN}(X)\p_m
X^M\p_n
X^N,\h C_{012}= c_{MNP}(X)\p_{0}X^{M}\p_{1}X^{N}\p_{2}X^{P}, \\
\nl &&\p_m=\p/\p\xi^m,\h m = (0,i) = (0,1,2),\\ \nl
&&(\xi^0,\xi^1,\xi^2)=(\tau,\sigma_1,\sigma_2),\h M =
(0,1,\ldots,10),\ea are the fields induced on the membrane
worldvolume from the background metric $g_{MN}$ and the background
3-form gauge field $c_{MNP}$, $\lambda^m$ are Lagrange
multipliers, $x^M=X^M(\xi)$ are the membrane embedding
coordinates, and $T_2$ is its tension. As shown in \cite{NPB656},
the above action is classically equivalent to the Nambu-Goto type
action \ba\nl S^{NG}= - T_2\int d^{3}\xi
\left(\sqrt{-\det{G_{mn}}}-\frac{1}{6}\varepsilon^{mnp}
\p_{m}X^{M}\p_n X^N \p_{p}X^{P} c_{MNP}\right)\ea and to the
Polyakov type action \ba\nl S^{P}= - \frac{T_2}{2}\int
d^{3}\xi\left[\sqrt{-\gamma}\left(\gamma^{mn} G_{mn}-1\right) -
\frac{1}{3}
\varepsilon^{mnp}\p_{m}X^{M}\p_nX^N\p_{p}X^{P}c_{MNP}\right],\ea
where $\gamma^{mn}$ is the auxiliary worldvolume metric and
$\gamma=\det\gamma_{mn}$. In addition, the action (\ref{oma})
gives a unified description for the tensile and tensionless
membranes.

The equations of motion for the Lagrange multipliers $\lambda^{m}$
generate the constraints \ba\label{00}
&&G_{00}-2\lambda^{j}G_{0j}+\lambda^{i}\lambda^{j}G_{ij}
+\left(2\lambda^0T_2\right)^2\det G_{ij}=0,\\
\label{0j} &&G_{0j}-\lambda^{i}G_{ij}=0.\ea

Further on, we will work in diagonal worldvolume gauge
$\lambda^{i}=0$, in which the action (\ref{oma}) and the
constraints (\ref{00}), (\ref{0j}) simplify to \ba\label{omagf}
&&S_{M}=\int d^{3}\xi \mathcal{L}_{M}= \int
d^{3}\xi\left\{\frac{1}{4\lambda^0}\Bigl[G_{00}-\left(2\lambda^0T_2\right)^2\det
G_{ij}\Bigr] + T_2 C_{012}\right\},
\\ \label{00gf} &&G_{00}+\left(2\lambda^0T_2\right)^2\det G_{ij}=0,
\\ \label{0igf} &&G_{0i}=0.\ea
Let us note that the action (\ref{omagf}) and the constraints
(\ref{00gf}), (\ref{0igf}) {\it coincide} with the usually used
gauge fixed Polyakov type action and constraints after the
following identification of the parameters $2\lambda^0T_2=L=const$
(see for instance \cite{27}).

Searching for membrane configurations in $AdS_4\times S^7$, which
correspond to the Neumann or Neumann-Rosochatius integrable
systems, we should first eliminate the membrane interaction with
the background 3-form field on $AdS_4$, to ensure more close
analogy with the strings on $AdS_5\times S^5$. To make our choice,
let us write down the background. It can be parameterized as
follows \ba\nl &&ds^2=(2l_p\mathcal{R})^2\left[-\cosh^2\rho
dt^2+d\rho^2+\sinh^2\rho\left(d\alpha^2+\sin^2\alpha
d\beta^2\right)+ 4d\Omega_7^2\right],
\\ \nl &&d\Omega_7^2= d\psi_1^2+\cos^2\psi_1
d\varphi_1^2\\ \nl &&+\sin^2\psi_1\left[d\psi_2^2+\cos^2\psi_2
d\varphi_2^2+ \sin^2\psi_2\left(d\psi_3^2+\cos^2\psi_3
d\varphi_3^2+\sin^2\psi_3 d\varphi_4^2\right)\right],
\\ \nl &&c_{(3)}=(2l_p\mathcal{R})^3\sinh^3\rho\sin\alpha dt\wedge
d\alpha\wedge d\beta.\ea Since we want the membrane to have
nonzero conserved energy and spin on $AdS$, one possible choice,
for which the interaction with the $c_{(3)}$ field disappears, is
to fix the angle $\alpha$: $\alpha=\alpha_0=const.$ The metric of
the corresponding subspace of $AdS_4$ is \ba\nl
ds^2_{sub}=(2l_p\mathcal{R})^2\left[-\cosh^2\rho
dt^2+d\rho^2+\sinh^2\rho d(\beta\sin\alpha_0)^2\right].\ea The
appropriate membrane embedding into $ds^2_{sub}$ and $S^7$ is
\ba\nl
&&Z_{\mu}=2l_p\mathcal{R}\mbox{r}_\mu(\xi^m)e^{i\phi_\mu(\xi^m)},\h
\mu=(0,1),\h \phi_{\mu}=(\phi_0,\phi_1)=(t,\beta\sin\alpha_0), \\
\nl &&W_{a}=4l_p\mathcal{R}r_a(\xi^m)e^{i\varphi_a(\xi^m)},\h
a=(1,2,3,4),\ea where $\mbox{r}_\mu$ and $r_a$ are real functions
of $\xi^m$, while $\phi_\mu$ and $\varphi_a$ are the isometric
coordinates on which the background metric does not depend. The
six complex coordinates $Z_{\mu}$, $W_{a}$ are restricted by the
two real embedding constraints \ba\nl
\eta^{\mu\nu}Z_{\mu}\bar{Z}_{\nu}+\left(2l_p\mathcal{R}\right)^2=0,\h
\eta^{\mu\nu}=(-1,1),\h
\delta_{ab}W_{a}\bar{W}_{b}-\left(4l_p\mathcal{R}\right)^2=0,\ea
or equivalently \ba\nl \eta^{\mu\nu} \mbox{r}_\mu\mbox{r}_\nu+1=0,
\h \delta_{ab} r_ar_b-1=0.\ea The coordinates $\mbox{r}_\mu$,
$r_a$ are connected to the initial coordinates, on which the
background depends, through the equalities \ba\nl
&&\mbox{r}_0=\cosh\rho,\h \mbox{r}_1=\sinh\rho,
\\ \nl &&r_1= \cos\psi_1,\h r_2= \sin\psi_1\cos\psi_2,
\\ \nl &&r_3= \sin\psi_1\sin\psi_2\cos\psi_3,\h
r_4= \sin\psi_1\sin\psi_2\sin\psi_3.\ea

For the embedding described above, the induced metric is given by
\ba\label{im} &&G_{mn}=\eta^{\mu\nu}\p_{(m}Z_\mu\p_{n)}\bar{Z_\nu}
+ \delta_{ab}\p_{(m}W_a\p_{n)}\bar{W_b}= \\ \nl
&&(2l_p\mathcal{R})^2\left[\sum_{\mu,\nu=0}^{1}\eta^{\mu\nu}
\left(\p_m\mbox{r}_\mu\p_n\mbox{r}_\nu +
\mbox{r}_\mu^2\p_m\phi_\mu\p_n\phi_\nu\right)+
4\sum_{a=1}^{4}\left(\p_mr_a\p_nr_a +
r_a^2\p_m\varphi_a\p_n\varphi_a\right)\right].\ea Correspondingly,
the membrane Lagrangian becomes \ba\nl
\mathcal{L}=\mathcal{L}_{M}+\Lambda_A(\eta^{\mu\nu}
\mbox{r}_\mu\mbox{r}_\nu+1)+\Lambda_S(\delta_{ab} r_ar_b-1),\ea
where $\Lambda_A$ and $\Lambda_S$ are Lagrange multipliers.

\subsection{Neumann and Neumann-Rosochatius integrable systems\\ from membranes}
\h Let us consider the following particular case of the above
membrane embedding \cite{MNR} \ba\label{nra}
&&Z_{0}=2l_p\mathcal{R}e^{i\kappa\tau}, Z_1=0,\h
W_{a}=4l_p\mathcal{R}r_a(\xi,\eta)e^{i\left[\omega_{a}\tau+\mu_a(\xi,\eta)\right]},
\\ \nl && \xi=\alpha\sigma_1+\beta\tau, \eta=\gamma\sigma_2+\delta\tau,\ea
which implies \ba\label{i} \mbox{r}_0=1,\h \mbox{r}_1=0,\h
\phi_0=t=\kappa\tau,\h
\varphi_a(\xi^m)=\varphi_a(\tau,\sigma_1,\sigma_2)=
\omega_{a}\tau+\mu_a(\xi,\eta).\ea Here $\kappa$, $\omega_{a}$,
$\alpha$, $\beta$, $\gamma$, $\delta$ are parameters, whereas
$r_a(\xi,\eta)$, $\mu_a(\xi,\eta)$ are arbitrary functions. As a
consequence, the embedding constraint $\eta^{\mu\nu}
\mbox{r}_\mu\mbox{r}_\nu+1=0$ is satisfied identically. For this
ansatz, the membrane Lagrangian takes the form ($\p_\xi=\p/\p\xi$,
$\p_\eta=\p/\p\eta$) \ba\nl
&&\mathcal{L}=-\frac{(4l_p\mathcal{R})^2}{4\lambda^0}
\left\{\left(8\lambda^0T_2l_p\mathcal{R}\alpha\gamma\right)^2
\sum_{a<b=1}^{4}\left[(\p_\xi r_a\p_\eta r_b-\p_\eta r_a\p_\xi r_b)^2\right. \right. \\
\nl &&+ \left. \left. (\p_\xi r_a\p_\eta\mu_b-\p_\eta
r_a\p_\xi\mu_b)^2r_b^2 + (\p_\xi\mu_a\p_\eta
r_b-\p_\eta\mu_a\p_\xi r_b)^2r_a^2\right.\right.
\\ \nl &&+\left.\left.(\p_\xi\mu_a\p_\eta\mu_b-\p_\eta\mu_a\p_\xi\mu_b)^2r_a^2r_b^2 \right]\right. \\
\nl &&+
\left.\sum_{a=1}^{4}\left[\left(8\lambda^0T_2l_p\mathcal{R}\alpha\gamma\right)^2
(\p_\xi r_a\p_\eta\mu_a-\p_\eta r_a\p_\xi\mu_a)^2-
\left(\beta\p_\xi\mu_a+\delta\p_\eta\mu_a+\omega_a\right)^2\right]r_a^2\right.
\\
\nl &&-\left.\sum_{a=1}^{4}\left(\beta\p_\xi r_a+\delta\p_\eta
r_a\right)^2+(\kappa/2)^2\right\}+\Lambda_S\left(\sum_{a=1}^{4}r_a^2-1\right).\ea
Now, we make the choice \ba\nl &&r_{1}=r_{1}(\xi),\h
r_{2}=r_{2}(\xi),\h \omega_3=\pm\omega_4=\omega,\\ \nl
&&r_3=r_3(\eta)=\epsilon\sin(b\eta+c),\h r_4=r_4(\eta)=\epsilon\cos(b\eta+c),\\
\nl &&\mu_1=\mu_1(\xi),\h \mu_2=\mu_2(\xi),\h
\mu_3,\mu_4=constants,\ea and receive (prime is used for $d/d\xi$)
\ba\nl &&\mathcal{L}=-\frac{(4l_p\mathcal{R})^2}{4\lambda^0}
\left\{\sum_{a=1}^{2}\left[(A^2-\beta^2)r_a'^2\right.+
\left.(A^2-\beta^2)r_a^2\left(\mu'_a-\frac{\beta\omega_a}{A^2-\beta^2}\right)^2
- \frac{A^2}{A^2-\beta^2}\omega_a^2 r_a^2\right]\right. \\
\nl &&+\left.
(\kappa/2)^2-\epsilon^2(\omega^2+b^2\delta^2)\right\} +
\Lambda_S\left[\sum_{a=1}^{2}r_a^2-(1-\epsilon^2)\right],\ea where
$ A^2\equiv \left(8\lambda^0T_2l_p\mathcal{R}\epsilon
b\alpha\gamma\right)^2$. A single time integration of the
equations of motion for $\mu_a$ following from the above
Lagrangian gives \ba\nl
\mu'_a=\frac{1}{A^2-\beta^2}\left(\frac{C_a}{r_a^2}+\beta\omega_a\right),\ea
where $C_a$ are arbitrary constants. Taking this into account, one
obtains the following effective Lagrangian for the coordinates
$r_a(\xi)$ \ba\nl &&L=\frac{(4l_p\mathcal{R})^2}{4\lambda^0}
\sum_{a=1}^{2}\left[(A^2-\beta^2)r_a'^2 -
\frac{1}{A^2-\beta^2}\frac{C_a^2}{r_a^2} -
\frac{A^2}{A^2-\beta^2}\omega_a^2 r_a^2\right]\\ \nl
&&+\Lambda_S\left[\sum_{a=1}^{2}r_a^2-(1-\epsilon^2)\right].\ea
This Lagrangian in full analogy with the string considerations
corresponds to particular case of the $n$-dimensional {\it
Neumann-Rosochatius integrable system}. For $C_a=0$ one obtains
{\it Neumann integrable system}, which describes two-dimensional
harmonic oscillator, constrained to remain on a circle of radius
$\sqrt{1-\epsilon^2}$.

Let us write down the three constraints (\ref{00gf}), (\ref{0igf})
for the present case. To achieve more close correspondence with
the string on $AdS_5\times S^5$, we want the third one,
$G_{02}=0$, to be satisfied identically. To this end, since
$G_{02}\sim (ab)^2\gamma\delta,$ we set $\delta=0$, i.e.
$\eta=\gamma\sigma_2$. Then, the first two constraints give \ba\nl
&&\sum_{a=1}^{2}\left[(A^2-\beta^2)r_a'^2+
\frac{1}{A^2-\beta^2}\frac{C_a^2}{r_a^2}+
\frac{A^2}{A^2-\beta^2}\omega_a^2
r_a^2\right]=\frac{A^2+\beta^2}{A^2-\beta^2}\left[(\kappa/2)^2-(\epsilon\omega)^2\right],
\\ \label{effcs}&&\sum_{a=1}^{2}\omega_{a}C_a +
\beta\left[(\kappa/2)^2-(\epsilon\omega)^2\right]=0.\ea

\subsection{Energy and angular momenta}
\h Due to the background isometries, there exist global conserved
charges. In our case, the background does not depend on $\phi_0=t$
and $\varphi_a$. Therefore, the corresponding conserved quantities
are the membrane energy $E$ and four angular momenta $J_a$, given
as spatial integrals of the conjugated to these coordinates
momentum densities \ba\nl E=-\int
d^2\sigma\frac{\p\mathcal{L}}{\p(\p_0 t)},\h J_a=\int
d^2\sigma\frac{\p\mathcal{L}}{\p(\p_0\varphi_a)},\h a=1,2,3,4.\ea
$E$ and $J_a$ can be computed by using the expression (\ref{im})
for the induced metric and the ansats (\ref{nra}), (\ref{i}).

In order to reproduce the string case, we can set $\omega=0$, and
thus $J_3=J_4=0$. The energy and the other two angular momenta are
given by \ba\nl
E=\frac{4\pi(l_p\mathcal{R})^2\kappa}{\lambda^0\alpha}\int d\xi,\h
J_a=\frac{\pi(4l_p\mathcal{R})^2}{\lambda^0\alpha(A^2-\beta^2)}\int
d\xi \left(\beta C_a + A^2\omega_a r_a^2\right),\h a=1,2.\ea From
here, by using the constraints (\ref{effcs}), one obtains the
energy-charge relation \ba\nl
\frac{4}{A^2-\beta^2}\left[A^2(1-\epsilon^2) +
\beta\sum_{a=1}^{2}\frac{C_a}{\omega_a}\right]\frac{E}{\kappa}
=\sum_{a=1}^{2}\frac{J_a}{\omega_a},\ea in full analogy with the
string case. Namely, for strings on $AdS_5\times S^5$, the result
in conformal gauge is \cite{NR2} \ba\nl
\frac{1}{\alpha^2-\beta^2}\left(\alpha^2+
\beta\sum_{a}\frac{C_a}{\omega_a}\right)\frac{E}{\kappa}
=\sum_{a}\frac{J_a}{\omega_a}.\ea

Let us point out that the membrane configuration considered here
corresponds exactly to the string embedding in the $R\times S^5$
subspace of $AdS_5\times S^5$ solution of type IIB string theory,
which is known to lead the Neumann and Neumann-Rosochatius
dynamical systems \cite{NR2}, including recently discovered giant
magnon and spiky string configurations.

\section{$SU(2)$ spin chain from membrane}
\h One of the predictions of AdS/CFT duality is that the string
theory on $AdS_5\times S^5$ should be dual to $\mathcal{N}=4$ SYM
theory in four dimensions. The spectrum of the string states and
of the operators in SYM should be the same. The first checks of
this conjecture {\it beyond} the supergravity approximation
revealed that there exist string configurations, whose energies in
the semiclassical limit are related to the anomalous dimensions of
certain gauge invariant operators in the planar SYM. On the field
theory side, it was found that the corresponding dilatation
operator is connected to the Hamiltonian of integrable Heisenberg
spin chain. On the other hand, it was established that there is
agreement at the level of actions between the continuous limit of
the $SU(2)$ spin chain arising in $\mathcal{N}=4$ SYM theory and a
certain limit of the string action in $AdS_5\times S^5$
background. Shortly after, it was shown that such equivalence also
holds for the $SU(3)$ and $SL(2)$ cases.

Here, we are interested in answering the question: is it possible
to reproduce this type of string/spin chain correspondence from
membranes on eleven dimensional curved backgrounds? It turns out
that the answer is positive at least for the case of M2-branes on
$AdS_4\times S^7$, as we will show below \cite{SCfromM}.

We will use our initial membrane embedding and fix \ba\nl
Z_{0}=2l_p\mathcal{R}e^{i\kappa\tau},\h Z_1=0,\ea which implies
$\mbox{r}_0=1$, $\mbox{r}_1=0$, $\phi_0=t=\kappa\tau$. Let us now
introduce new coordinates by setting \ba\nl
(\varphi_1,\varphi_2,\varphi_3,\varphi_4)=
\left(\frac{\kappa}{2}\tau+\alpha+\varphi,\frac{\kappa}{2}\tau+\alpha-\varphi,
\frac{\kappa}{2}\tau+\alpha+\phi,
\frac{\kappa}{2}\tau+\alpha-\tilde{\phi}\right)\ea and take the
limit $\kappa\to\infty$, $\p_0\to 0$, $\kappa\p_0$ - finite. In
this limit, we obtain the following expression for the membrane
Lagrangian \ba\nl
\mathcal{L}&=&\frac{(2l_p\mathcal{R})^2}{\lambda^0}\kappa\left(\p_0\alpha
+
\sum_{k=1}^{3}\nu_k\p_0\rho_k\right)-\lambda^0T_2^2(4l_p\mathcal{R})^4
\left\{\sum_{a<b=1}^{4}(\p_1r_a\p_2r_b-\p_2r_a\p_1r_b)^2\right. \\
\nl &+&
\left.\sum_{a=1}^{4}\sum_{k=1}^{3}\mu_k(\p_1r_a\p_2\rho_k-\p_2r_a\p_1\rho_k)^2-
\sum_{a=1}^{4}\left(\p_1r_a\sum_{k=1}^{3}\nu_k\p_2\rho_k
- \p_2r_a\sum_{k=1}^{3}\nu_k\p_1\rho_k\right)^2\right. \\ \nl &+&
\left.\sum_{k<n=1}^{3}\mu_k\mu_n(\p_1\rho_k\p_2\rho_n-\p_2\rho_k\p_1\rho_n)^2\right.\\
\nl
&-&\left.\sum_{k=1}^{3}\mu_k\left(\p_1\rho_k\sum_{n=1}^{3}\nu_n\p_2\rho_n
- \p_2\rho_k\sum_{n=1}^{3}\nu_n\p_1\rho_n\right)^2\right\}+
\Lambda_S\left(\sum_{a=1}^{4}r_a^2-1\right),\ea where \ba\nl
(\mu_1,\mu_2,\mu_3)=(r_1^2+r_2^2,r_3^2,r_4^2),
(\nu_1,\nu_2,\nu_3)=(r_1^2-r_2^2,r_3^2,-r_4^2),
(\rho_1,\rho_2,\rho_3)=(\varphi,\phi,\tilde{\phi}) .\ea

Now, we are ready to face our main problem: how to reduce this
Lagrangian to the one corresponding to the thermodynamic limit of
spin chain, {\it without shrinking the membrane to string}? We
propose the following solution of this task: \ba\nl &&
\alpha=\alpha(\tau,\s_1),\h r_1=r_1(\tau,\s_1),\h
r_2=r_2(\tau,\s_1),\\ \nl
&&r_3=r_3(\tau,\s_2)=\epsilon\sin[b\s_2+c(\tau)],\h
r_4=r_4(\tau,\s_2)=\epsilon\cos[b\s_2+c(\tau)],\\ \nl
&&\varphi=\varphi(\tau,\s_1),\h
\epsilon,b,\phi,\tilde{\phi}=constants,\h \epsilon^2<1. \ea These
restrictions lead to \ba\nl \mathcal{L}
&=&\frac{(2l_p\mathcal{R})^2}{\lambda^0}\kappa\left[\p_0\alpha +
(r_1^2-r_2^2)\p_0\varphi\right]-\lambda^0(\epsilon
bT_2)^2(4l_p\mathcal{R})^4
\left\{\sum_{a=1}^{2}(\p_1r_a)^2\right.\\ \nl &+&
\left.\left[(r_1^2+r_2^2)-(r_1^2-r_2^2)^2\right](\p_1\varphi)^2\right\}
+\Lambda_S\left[\sum_{a=1}^{2}r_a^2-(1-\epsilon^2)\right].\ea If
we introduce the parametrization \ba\nl
r_1=(1-\epsilon^2)^{1/2}\cos\psi,\h
r_2=(1-\epsilon^2)^{1/2}\sin\psi,\ea the new variable
$\tilde{\alpha}$=$\alpha/(1-\epsilon^2)$, and take the limit
$\epsilon^2\to 0$ neglecting the therms of order higher than
$\epsilon^2$, we will receive \ba\nl
\frac{\mathcal{L}}{1-\epsilon^2}=\frac{(2l_p\mathcal{R})^2}{\lambda^0}\kappa\left[\p_0\tilde{\alpha}
+\cos(2\psi)\p_0\varphi\right] -\lambda^0(\epsilon
bT_2)^2(4l_p\mathcal{R})^4\left[(\p_1\psi)^2
+\sin^2(2\psi)(\p_1\varphi)^2\right].\ea As for the membrane
action corresponding to the above Lagrangian, it can be
represented in the form \ba\nl S_M=\frac{\mathcal{J}}{2\pi}\int dt
d\s\left[\p_t\tilde{\alpha} +
\cos(2\psi)\p_t\varphi\right]-\frac{\tilde{\lambda}}{4\pi\mathcal{J}}\int
dt d\s
\left[\left(\p_\s\psi\right)^2+\sin^2(2\psi)(\p_\s\varphi)^2\right],\ea
where $\mathcal{J}$ is the angular momentum conjugated to
$\tilde{\alpha}$, $t=\kappa\tau$ and \ba\nl
\tilde{\lambda}=2^{15}(\pi^2\epsilon bT_2)^2(l_p\mathcal{R})^6.\ea
This action corresponds to the thermodynamic limit of $SU(2)$ {\it
integrable spin chain} \cite{K0311}.

\vspace*{.5cm}

{\bf Acknowledgments} \vspace*{.2cm}

This work is supported by NSFB grants $F-1412/04$ and
$VU-F-201/06$.

\end{document}